\documentclass[apj,twocolumn,floatfix,nofootinbib,trackchanges,table,linenumbers]{openjournal}
\usepackage[breaklinks,colorlinks,citecolor=blue,urlcolor=blue,linkcolor=blue]{hyperref}

\usepackage{physics}
\usepackage{graphicx}
\usepackage{microtype}
\usepackage{xspace}
\usepackage{amsmath}
\usepackage{booktabs}
\usepackage{multirow}
\usepackage{xcolor}
\usepackage[normalem]{ulem}
\usepackage[dvipsnames]{xcolor} 
\definecolor{ultra}{HTML}{0B6572}
\definecolor{udark}{HTML}{0A5681}
\renewcommand{\arraystretch}{1.2}  
\usepackage{listings}
\usepackage{orcidlink}
\allowdisplaybreaks

\newcommand{\rev}[1]{\textbf{}}
\newcommand{\ff}{\ensuremath{\mathcal{FF}}\xspace}
\newcommand{\rhonetobs}{\ensuremath{\hat{\rho}^\mathrm{net}_\mathrm{obs, \phi}}\xspace}

\begin{document}

\title{Semianalytic Sensitivity Estimates for Out-of-Bank Gravitational-Wave Signals} 
\shortauthors{Vijaykumar \& Essick}
\shorttitle{Semianalytic Sensitivity Estimates for Out-of-Bank Gravitational-Wave Signals}
\author{\vspace{-4em} Aditya Vijaykumar\,\orcidlink{0000-0002-4103-0666}$^{1}$}
\author{Reed Essick\,\orcidlink{0000-0001-8196-9267}$^{1,2,3}$}

\affiliation{$^1$Canadian Institute for Theoretical Astrophysics, University of Toronto, 60 St George St,  Toronto, ON M5S 3H8, Canada}
\affiliation{$^2$Department of Physics, University of Toronto, 60 St. George Street, Toronto, ON M5S 1A7}
\affiliation{$^3$David A. Dunlap Department of Astronomy, University of Toronto, 50 St. George Street, Toronto, ON M5S 3H4}

\begin{abstract}
    Estimating the sensitivity of gravitational-wave searches is important for a wide variety of scientific applications spanning astrophysics and fundamental physics. 
    In this work, we develop a fast semianalytic approximation for estimating matched-filter search sensitivity to physical effects not explicitly modeled in the template bank.
    This approximation utilizes fitting factors, i.e., the maximum overlap of a candidate signal  over the search bank. 
    As illustrations, we compare our estimates to the actual performance of  searches against spinning binary neutron stars, and evaluate search sensitivity to compact binaries possessing orbital eccentricity or deviations from general relativity.
    Our work thus paves the way for fast sensitivity estimates for a variety of applications, including unmodeled effects in template banks such as deviations from general relativity, environmental effects, gravitational lensing, and waveform/calibration systematics.
\end{abstract}

\section{Introduction}
\label{sec:introcution}

\begin{figure}[!htbp]
    \centering
    \includegraphics[scale=1]{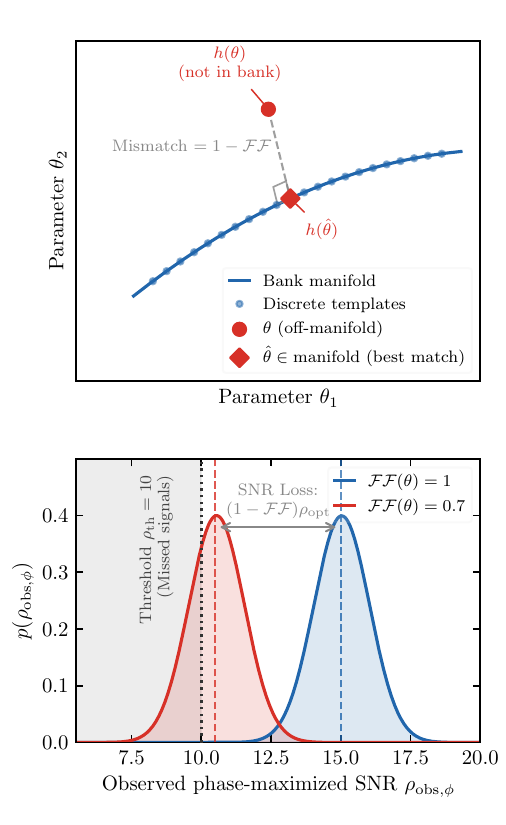}
    \caption{
        \textit{Illustration of our method}. The top panel shows the template bank manifold in the blue solid line, with the discreteness of the bank denoted by blue points. A candidate signal described by $h(\theta)$ and optimal SNR $\rho_\mathrm{opt}(\theta)$ outside the bank has a best match template $h(\hat{\theta})$ in the bank, incurring a mismatch $1 - \ff(\theta)$ in the process. Thus, in the bottom panel, this translates to a reduced non-centrality parameter $\ff(\theta) \rho_\mathrm{opt}$ for the distribution of the observed phase-maximized SNR $\rho_\mathrm{obs, \phi}$, reducing the detectability of the signal. In this setup, $\ff(\theta) = 0.7$ and $29\%$ signals are missed, whereas if $\ff(\theta) = 1$ only $2.3 \times 10^{-5}\%$ signals are missed.
    }
    \label{fig:illustration}
\end{figure}

One distinctive feature of  gravitational wave (GW) astronomy, as opposed to electromagnetic (EM) astronomy, is that selection effects can be estimated with high precision.
This follows directly from  the fact that the GW signal from compact binaries is completely specified by a fixed set of source parameters with very high precision, and the GW signal is not affected by matter along its propagation path, modulo gravitational lensing.
This feature of GW signals is crucial for correcting selection biases and obtaining unbiased estimates of the underlying source population of merging binaries.

In practice, the selection function is estimated by building large sets of simulated GW signals embedded in realistic detector noise, referred to as \textit{injections}. 
These injections are processed by search pipelines exactly as they would process real data; for each injection,  detection statistics such as the signal-to-noise ratio (SNR), false alarm rate (FAR), and the probability of astrophysical origin ($p_{\rm astro}$) are recorded~\citep{2026arXiv260527224T, 2023ApJ...946...59N, 2025arXiv250117939M}. 
A realistic injection set for current catalogs~\citep{2026arXiv260527225T, 2025arXiv250818082T} typically contains ${\sim} 10^6$ injections~\citep{2025PhRvD.112j2001E, ligo_scientific_collaboration_2025_16740128}, making this process significantly intensive in terms of computational- and person-power.
The required number of injections also grows approximately linearly with catalog size~\citep{2022arXiv220400461E}.
Therefore, while in-principle straightforward, calculating the selection function is nontrivial in practice.
These difficulties compound if selection effects need to be estimated for a wide variety of applications, e.g., for eccentric sources, for signals motivated by modified gravity theories, etc. Approaches that either learn the selection function from injection sets~\citep{2020PhRvD.102j3020G, 2020arXiv201201317T,2024PhRvD.110l3041C} or calibrate simple-to-evaluate expression to them~\citep{2024PhRvD.109f3013M, 2025CQGra..42d5008L} also suffer from related issues because they are predicated on the existence of large injection sets for training data.

Thus, methods for making this process faster are desirable.
Semianalytic approaches~\citep{1993PhRvD..47.2198F, 2023PhRvD.108d3011E} are one solution.
The fundamental idea behind this proposal is that the observed SNR is distributed around the optimal SNR following a non-central $\chi$ distribution with $2N_{\rm det}$ degrees of freedom, where $N_{\rm det}$ is the total number of detectors in the network. The spread around the optimal SNR is driven by fluctuations in the detector noise. 
\cite{2023PhRvD.108d3011E} showed that a semianalytic SNR threshold of ${\approx}10 $ reproduces the detected distribution of the O3 injection set under a ${\rm FAR} < 1 \, {\rm yr}^{-1}$ cutoff.
Similar ideas have been explored in other works~\citep{Wysocki2019,2024CQGra..41l5002G}. 

While this method is effective, it suffers from a few drawbacks. 
First, current matched-filter searches used by the LIGO-Virgo-KAGRA (LVK) Collaboration~\citep{2015CQGra..32g4001L,2015CQGra..32b4001A,2021PTEP.2021eA101A} do not include the effect of eccentricity or spin-precession in their template banks.
This means that traditional semianalytic estimates, which assume all the relevant physics is included within the template bank, only provide an upper bound on sensitivity to such effects.
Secondly, the method also does not account for the finite spacing of template banks and assumes that the template bank is infinitely dense in all regions of the signal parameter space.
As evidenced by choices in, e.g.,~\citet{2024PhRvD.109d4066S} and discussed in \S\ref{sec:applications}, real template banks are discrete and might lack coverage in some regions of the source parameter space.
As an example, consider an extreme case where a signal has zero overlap with every template in a bank. Because the semianalytic prescription is independent of template bank structure and coverage, it would categorize such a signal as detectable based solely on its true SNR.
In reality, however, the search sensitivity is identically zero.
Additionally, while traditional semianalytic estimates might be able to mimic the detection statistic, they cannot provide other information that searches provide, e.g., the best match template parameters for each injection.

In this work, we develop a framework for semianalytic sensitivity estimates that addresses these issues.
Instead of assuming that a signal always has a closely matching template in the bank, we explicitly maximize the overlap of a candidate signal across the finite template bank before projecting it onto a detector network and incorporating the effects of noise.
We demonstrate that our method reproduces a range of results reported in the literature.
Furthermore, thanks to a GPU-accelerated maximization procedure, our technique is highly efficient and can produce realistic sensitivity estimates within just a few hours.

This paper is structured as follows.
In \S\ref{sec:methods}, we lay out the mathematical framework underlying our method and also detail various analysis settings and choices.
In \S\ref{sec:applications}, we concentrate on three different applications:  gaps and discreteness in existing template banks (\S\ref{sec:incomplete banks}),  missing physics, like eccentricity, that is known but not included in the search manifold (\S\ref{sec:known missing physics}), and  missing physics that is unknown, like deviations from general relativity, and is therefore also not included in the search manifold (\S\ref{sec:unknown missing physics}).
We discuss other potential avenues for applications in \S\ref{sec:discussion} along with caveats and possible future improvements.

\section{Methods}
\label{sec:methods}


\subsection{Nomenclature}
\label{sec:nomenclature}

We model GW data in the frequency-domain $d(f)$ as
\begin{equation}
    d(f) = h(f;\theta) + n(f),
\end{equation}
where $h(f;\theta)$ represents the GW signal characterized by the parameters $\theta$ and $n(f)$ is additive, zero-mean, stationary Gaussian noise with one-sided power spectral density (PSD) denoted by $S_n(f)$.
We define the noise-weighted inner product in the frequency domain as
\begin{equation}
    \ip{a}{b} = 4 \int_{f_{\text{low}}}^{f_{\text{high}}} \frac{a(f)\, b^*(f)}{S_n(f)}\, df,
\end{equation}
where $b^*(f)$ is the complex conjugate of $b(f)$.
Note that $\ip{a}{b}$ is, in general, complex.

To identify potential signals, we filter the data against a template bank $\{h(\theta_T)\}$.
By maximizing the stationary Gaussian likelihood with respect to the signal's amplitude, we can derive the matched-filter response of the data against a single template $h(\theta_T)$
\begin{equation}
    \hat{\rho}(\theta_T) = \frac{\ip{d}{h(\theta_T)}}{\rho_\mathrm{opt}(\theta_T)},
\end{equation}
where $\rho_\mathrm{opt}(\theta_T) = \sqrt{\ip{h(\theta_T)}{h(\theta_T)}}$ is the optimal SNR of the template.
Writing $\hat{\rho} = \hat{\rho}_R + i\hat{\rho}_I$, the real and imaginary parts are independent Gaussian random variables with unit variance.
If the data contains a signal described by $\theta$, then the expected values of the real and imaginary filter responses are
\begin{align}
    \label{eq:expval-1}
    \mathbb{E}[\tilde{\rho}_R] &= \mathfrak{M}(\theta_T;\theta)\,\rho_\mathrm{opt}(\theta)\cos\Delta\phi, \\
     \label{eq:expval-2}
    \mathbb{E}[\tilde{\rho}_I] &= \mathfrak{M}(\theta_T;\theta)\,\rho_\mathrm{opt}(\theta)\sin\Delta\phi,
\end{align}
where
\begin{equation}
    \mathfrak{M}(\theta_T;\theta) = \frac{\abs{\ip{h(\theta)}{h(\theta_T)}}}{\rho_\mathrm{opt}(\theta_T)\,\rho_\mathrm{opt}(\theta)}
    \label{eq:match}
\end{equation}
is the match between the signal and template and $\Delta\phi$ is their relative phase offset.
For compact binary coalescences (CBCs), $\Delta \phi$ could correspond to the orbital phase at coalescence.
The phase-maximized filter response is then
\begin{equation}
    \hat{\rho}_{\phi}(\theta_T;\theta) = \sqrt{\hat{\rho}_R^2(\theta_T;\theta) + \hat{\rho}_I^2(\theta_T;\theta)}.
\end{equation}
One can show that $\hat{\rho}_{\phi}$ follows a non-central $\chi$ distribution with two degrees of freedom and non-centrality parameter 
\begin{equation}
    \lambda = \mathfrak{M}(\theta_T;\theta)\rho_\mathrm{opt}(\theta).
\end{equation}

Another quantity that is useful to calculate is the \textit{fitting factor}~\citep{1995PhRvD..52..605A}, which is the match in Equation~\eqref{eq:match} maximized over the template bank
\begin{equation}
    \ff(\theta) = \max_{\theta_T \in \mathrm{bank}} \mathfrak{M}(\theta_T;\theta).
\end{equation}
By definition, $0 \le \ff(\theta) \le 1$. The equality $\ff(\theta) = 1$ is  satisfied when the best-matching template corresponds to the true signal.

In a network of $N_\mathrm{det}$ detectors, the network phase-maximized filter response is obtained by summing the individual detector responses in quadrature (i.e., maximizing over $\Delta \phi$ separately in each detector)
\begin{equation}
    \hat{\rho}^\mathrm{net}_{\phi}(\theta_T;\theta) = \sqrt{\sum_{i=1}^{N_{\mathrm{det}}} \left[\hat{\rho}^i_{\phi}(\theta_T;\theta)\right]^2}.
    \label{eq:network-mf}
\end{equation}
$\hat{\rho}^\mathrm{net}_\phi(\theta_T)$ is non-central $\chi$ distributed with $2 N_\mathrm{det}$ degrees of freedom and non-centrality parameter
\begin{equation}
    \lambda^\mathrm{net} = \sqrt{\sum\limits_i^{N_\mathrm{det}} \left[\mathfrak{M}^i \rho_\mathrm{opt}^i \right]^2}
\end{equation}
Importantly, the optimal SNR ($\rho^i_\mathrm{opt}$), match ($\mathfrak{M}^i$), and fitting factor ($\ff^i$) can be different in each detector because the PSDs can be different in each detector.

Searches often maximize the network response over the template bank, and we define
\begin{equation}
    \rhonetobs(\theta) = \max_{\theta_T \in \mathrm{bank}} \left\{ \hat{\rho}^\mathrm{net}_{\phi}(\theta_T;\theta) \right\}.
    \label{eq:rhonetobs}
\end{equation}


\subsection{Approximating search behavior}
\label{sec:approximating search behavior}

In most matched-filter searches, the incoming data are filtered against a template bank and \rhonetobs from  Eq.~\eqref{eq:rhonetobs} is reported as one detection statistic.
However, one can attempt to approximate \rhonetobs in a semianalytic fashion following the distribution of the matched-filter responses from \S\ref{sec:nomenclature}.
In particular, one reasonable approximation could be $\hat{\rho}_{\phi}(\theta_T;\theta) \approx \hat{\rho}_{\phi}(\theta;\theta)$, in which case
\begin{equation}
    \rhonetobs(\theta) \approx \hat{\rho}^\mathrm{net}_{\phi}(\theta;\theta) = \sqrt{\sum_{i=1}^{N_{\mathrm{det}}} \qty[\hat{\rho}^i_{\phi}(\theta;\theta)]^2}.
    \label{eq:essick-approx}
\end{equation}
Since the RHS of Equation~\eqref{eq:essick-approx} is the quadrature sum of $2N_\mathrm{det}$ independent Gaussian random variables, $\rhonetobs(\theta)$ also follows a non-central $\chi$ distribution with $2N_\mathrm{det}$ degrees of freedom and non-centrality parameter
\begin{equation}
    \lambda = \rho_\mathrm{opt}^\mathrm{net}(\theta) = \sqrt{\sum_i^{N_\mathrm{det}} \qty[\rho_\mathrm{opt}^i(\theta)]^2}.
    \label{eq:essick-approx-2}
\end{equation}
\citet{2023PhRvD.108d3011E} shows that this approximation reproduces the behavior of searches under some choice of a threshold $\rhonetobs\ge\rho_\mathrm{thr}$. The effectiveness of this approximation is not surprising. Ideal search template banks are constructed to ensure $\ff(\theta)\approx1$ across the parameter space~\citep{1996PhRvD..53.6749O}; with this, the expectation value of the phase-maximized matched filter response approaches the optimal SNR. 
\begin{figure*}
    \centering
    \includegraphics[scale=1]{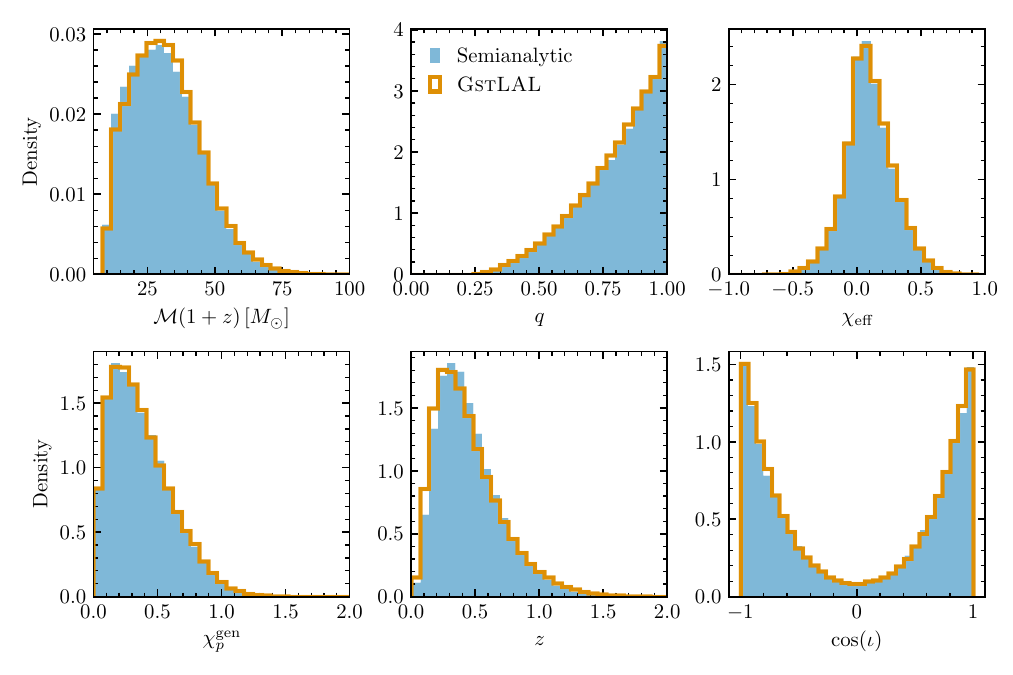}
    \caption{\emph{Distributions of detected source parameters from the  \textsc{GstLAL} results and the semianalytic prescription}. We assume a a population model following Equation~\eqref{eq:popmodel} with $m_\mathrm{min} = 10 \, M_\odot$ and $m_\mathrm{max} = 40\,M_\odot$. The distributions match between the semianalytic prescription and the real \textsc{GstLAL}-inspiral search results.}
    \label{fig:detected-dist}
    \vspace{0.75cm}
\end{figure*}

\begin{figure*}
    \centering
    \includegraphics[scale=1]{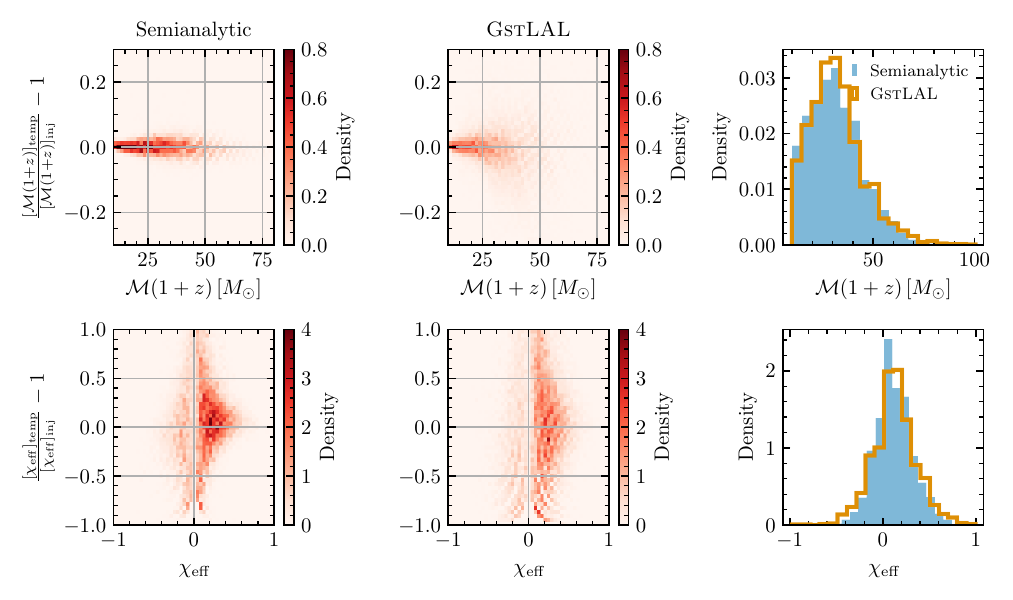}
    \caption{
        \emph{Distribution of best-match template parameters for $m_\mathrm{min} = 10 \, M_\odot$ and $m_\mathrm{max} = 40 \, M_\odot$}. 
        In the first row, the first panel  shows the histogram of the injected $\mathcal{M}(1+z)$ (horizontal axis) and the fractional error (vertical axis) for the semianalytic estimates. The `inj' and `temp' subscripts refer to the injected value and the corresponding recovered template value respectively. 
        The second panel shows the same quantities as the first panel but for the actual \textsc{GstLAL} search results.
        The third panel compares the distribution of recovered templates in our semianalytic estimates and from \textsc{GstLAL} search results.
        All the results have similar trends between the semianalytic and \textsc{GstLAL} results, with differences increasing towards high mass.
    } 
    \label{fig:lowmass-scatter}
\end{figure*}

\begin{figure*}
    \centering
    \includegraphics[scale=1]{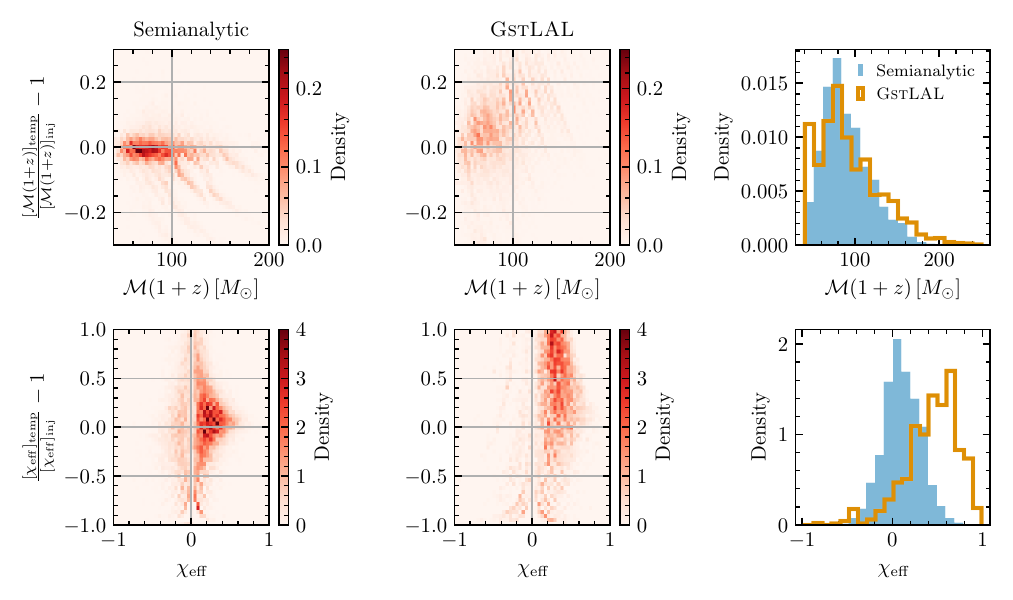}
    \caption{
        \textit{Distribution of best-match template parameters for $m_\mathrm{min} = 40 \, M_\odot$ and $m_\mathrm{max} = 100 \, M_\odot$}.
        The panels describe the same quantities as those in Figure~\ref{fig:lowmass-scatter}.
        However, we see significant differences in the template parameters, especially for $\chi_\mathrm{eff}$.
    }
    \label{fig:highmass-scatter}
\end{figure*}

However, when considering realistic, incomplete banks or unmodeled physical effects, we will generically have $\ff(\theta) < 1$.
To account for these effects when estimating sensitivity, it is then essential to include $\ff(\theta)$ in our estimates---the recovered search SNRs will be reduced due to the $\ff(\theta) <1 $ leading to a loss of detections, thus impacting the sensitivity of GW searches.
Thus, we approximate the expected value of the filter response as $\mathbb{E}[\hat{\rho}_R^i] \ff^i(
\theta) $.
To reduce computational load, we further assume that the template yielding the maximum match and its corresponding $\ff(\theta)$ are identical across all detectors.
Strictly speaking, this only makes sense if the detectors have identical PSDs, which is approximately true for the two LIGO detectors but not true in general.
This simplifies the non-centrality parameter
\begin{equation}
    \lambda = \ff(\theta) \rho_\mathrm{opt}^\mathrm{net}(\theta). \label{eq:THIS ONE}
\end{equation}
Future iterations of our prescription could easily relax this assumption.

Figure~\ref{fig:illustration} illustrates our procedure, which proceeds as follows. For each injection, we first compute the fitting factor $\ff(\theta)$ for a given template bank. We then evaluate the optimal network SNR, $\rho_\mathrm{opt}^\mathrm{net}(\theta)$, and model the observed network SNR, \rhonetobs, by drawing from a $\chi$-distribution with $2N_\mathrm{det}$ degrees of freedom and a non-centrality parameter $\lambda = \ff(\theta) \rho^\mathrm{net}_\mathrm{opt}(\theta)$. This methodology deviates from actual search pipelines, which perform the maximization in the right-hand side of Equation~\eqref{eq:rhonetobs} directly on the \emph{noisy data}. In contrast, we maximize the signal response over the template bank prior to incorporating noise. These operations do not commute—specific noise realizations can shift the maximum SNR to a template distinct from the true best-match template. Nevertheless, we demonstrate in \S\ref{sec:applications} that this approximation remains robust and accurate for the majority of scenarios and applications considered in this work.

For each injection, our prescription estimates $\rhonetobs$ by calculating $\ff(\theta)$ using a single reference PSD and separately computing $\rho^i_\mathrm{opt}(\theta)$ for each detector to obtain the non-centrality parameter in Eq.~\ref{eq:THIS ONE}.
The former entails calculating a vast number of inner products, which traditionally poses a computational bottleneck ($\gtrsim 10^{12}$ inner products for a standard injection set and template bank; \citealt{2025PhRvD.112j2001E}).
We overcome this limitation by employing GPU-compatible waveforms, implemented in the \texttt{ripple} software package~\citep{Edwards2024} via the \texttt{jax} array backend. 

For the analyses presented in \S\ref{sec:applications}, we find that the time $T$ required to compute $\ff(\theta)$ on a single \textsc{Nvidia Titan V} GPU scales as:
\begin{equation}
    \frac{T}{1 \, \mathrm{hr}} \approx \left( \frac{N_\mathrm{inj}}{1.5 \times 10^4} \right) \left( \frac{N_\mathrm{temp}}{10^6} \right) \left( \frac{\Delta f}{2 \, \mathrm{Hz}} \right)
\end{equation}
where $N_\mathrm{inj}$ is the number of injections, $N_\mathrm{temp}$ is the number of templates, and $\Delta f$ is the frequency grid spacing used to generate both the injections and templates (see also Appendix~\ref{appendix:Delta f}).
Once \rhonetobs is calculated for each injection, the only free parameter remaining to define detectability is the threshold $\rho_\mathrm{thr}$ placed on \rhonetobs.
It is possible to calibrate $\rho_\mathrm{thr}$ by comparing to real search results if they are available. For instance, 
\citet{2023PhRvD.108d3011E} found $\rho_\mathrm{thr}=10$ by comparing semianalytic estimates to search estimates in O3~\citep{2023PhRvX..13a1048A, ligo_scientific_collaboration_and_virgo_2023_7890398}.
We fix  $\rho_\mathrm{thr} = 10$ throughout the paper, but see Appendix~\ref{appendix:mass-dependent-rhoth} for additional explorations in calibrating $\rho_\mathrm{thr}$. 


\section{Applications}
\label{sec:applications}
In this section, we demonstrate the utility of our method by applying it to scenarios involving incomplete and finite template banks when all physics is known and included  (\S\ref{sec:incomplete banks}), signals for which known physical effects are not included in the template bank (\S\ref{sec:known missing physics}), and signals for which the full physical description is not necessarily known (\S\ref{sec:unknown missing physics}). We use $\mathcal{M}$ to denote chirp mass in the source frame, $q$ to denote mass ratio, $m_1$ and $m_2$ to denote primary and secondary component masses in the source frame, $M$ to denote total mass in the source frame,  $D_L$ to denote luminosity distance, $z$ to denote redshift, $\chi_\mathrm{eff}$ to denote the effective inspiral spin, $\iota$ to denote inclination of the orbit, $\chi_p^\mathrm{gen}$ to denote generalized effective precessing spin~\citep{2021PhRvD.103f4067G}, and $e$ to denote eccentricity.


\subsection{Discrete and Incomplete Template Banks}
\label{sec:incomplete banks}
We first show that our semianalytic prescriptions can faithfully reproduce search behavior in the simplest case of discrete (but approximately complete) template banks. For this purpose, we construct a semianalytic injection set for BBHs, and compare it to \textsc{GstLAL} sensitivity estimates from the first part of the LVK's fourth observing run (O4a)~\citep{2025ApJ...995L..18A}. 
The injection set that the sensitivity estimates were run on as well as our semianalytic injection set assume $\textsc{IMRPhenomXPHM}$~\citep{2021PhRvD.103j4056P} as the waveform approximant, and this approximant includes the effects of precession as well as higher modes on the signal. However, the template bank of \textsc{GstLAL} and other searches in O4~\citep{2012PhRvD..85h1504C, 2017PhRvD..95d2001M,
2017arXiv171108743R,
2017PhRvD..95j4045R,
2020PhRvD.101b2003H, 2021SoftX..1400680C, 2024PhRvD.109d4066S, 2023arXiv230607190R, 2023PhRvD.108d3004T, 2024PhRvD.109d2008E, 2025arXiv250606497J, 2025arXiv250523959J,2025CQGra..42j5009A,2021ApJ...923..254D,2022PhRvD.105b4023C} assume aligned-spin signals. While this does lead to some loss in sensitivity, it has been shown that the loss is small for most regions in the parameter space~\citep{2016PhRvD..94b4012H,2018PhRvD..97b3004H}. In any case, our semianalytic prescription can pick up on these differences if they were to be large.

To compare our semianalytic injection set to the one produced for O4a, we reweigh all our results to a fiducial mass model given by $p(m_1, m_2 | m_\mathrm{min}, m_\mathrm{max}) = p(m_1 | m_\mathrm{min}, m_\mathrm{max}) p(m_2 | m_1 , m_\mathrm{min}, m_\mathrm{max})$ where
\onecolumngrid
\begin{align}
    p(m_1 | m_\mathrm{min}, m_\mathrm{max})&= \frac{m_1^{-1}}{\ln(\dfrac{m_{\max}}{m_{\min}})}  \ ; \ m_\mathrm{min} \leq m_1 \leq m_\mathrm{max} \nonumber \\
    &= 0 \ ; \ \text{otherwise} \nonumber \\
    p(m_2 | m_1 ,m_\mathrm{min}, m_\mathrm{max}) &= \frac{2m_2}{m_1^2 - m_{\min}^2} \Theta(m_\mathrm{min} \leq m_2 \leq m_1).
    \label{eq:popmodel}
\end{align}
\twocolumngrid
The distribution of all other parameters is held fixed to the fiducial choices made in the injection set.
In Figure~\ref{fig:detected-dist}, we compare the detected distributions under Equation~\ref{eq:popmodel} with parameters $m_\mathrm{min}=10\,M_\odot$ and $m_\mathrm{max}=40\,M_\odot$. We find that the detected distribution agree, similar to the results of \citet{2023PhRvD.108d3011E}.

A useful byproduct of the $\ff(\theta)$ calculation is an estimate of the best-match template parameters. In Figures~\ref{fig:lowmass-scatter} and \ref{fig:highmass-scatter}, we compare the recovered templates from the actual \textsc{GstLAL} search to our semianalytic approximation for two injected populations: one with $m_\mathrm{min}=10\,M_\odot$ and $m_\mathrm{max}=40\,M_\odot$, and another with $m_\mathrm{min}=40\,M_\odot$ and $ m_\mathrm{max}=100\,M_\odot$.
We find that the distributions of $\mathcal{M}(1+z)$, and the mass ratio, $q$, for the best-match templates show good agreement between our semianalytic prescription and the real \textsc{GstLAL} injections. However, while the effective spin ($\chi_\mathrm{eff}$) distributions of the best-match templates align closely for the low-mass population, they diverge significantly for the high-mass population. While we did not investigate the reason for the discrepancy in detail, it is likely to be related to other operations in \textsc{GstLAL} that we do not emulate e.g., data conditioning, signal consistency tests. For instance,  non-Gaussian noise artifacts could preferentially respond to the shorter negative $\chi_\mathrm{eff}$ templates, substantially enhancing the noise background at those templates and shifting the preferred event to positive $\chi_\mathrm{eff}$ values. Nevertheless, these estimates of template parameters may be  valuable; for instance, they can be utilized in population analyses that apply parameter cuts to probe specific features in the binary black hole mass distribution~\citep{2025CQGra..42v5008K}, or to mimic mass thresholds used for selecting events in certain tests of general relativity~\citep{2026arXiv260319020T}.

\begin{figure}[!tbp]
    \centering
    \includegraphics{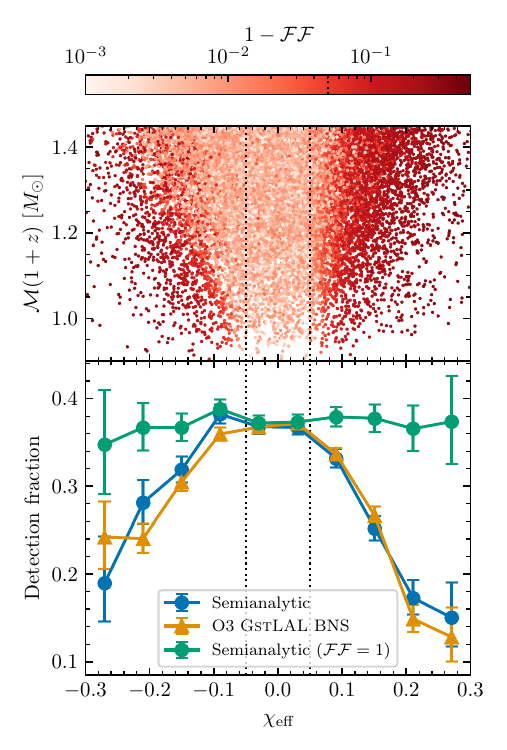}
    \caption{
        \textit{Gaps in the template bank}.
        The top panel shows the $1 -\ff$ values for BNS injections in O3 with $\mathcal{M}(1+z) < 1.45$.
        While \ff is large for points with $\abs{\chi_\mathrm{eff}} < 0.05$, it drops sharply outside this range.
        This is a direct consequence of the lack of templates with $\abs{\chi_\mathrm{eff}} > 0.05 $ points in the \textsc{GstLAL} template bank.
        In the bottom panel, we show the detection fraction of these injections as a function of $\chi_\mathrm{eff}$ also comparing it to results from the \textsc{GstLAL} search in O3.  
        The detection fraction drops steeply beyond $\abs{{\chi}_\mathrm{eff}} > 0.05$, dropping by a factor of $>3$ at $\chi_\mathrm{eff}=0.3$.
        The semianalytic sensitivity curves (orange) tracks the real search results (blue) within their uncertainty regions.
        Semianalytic sensitivity estimates that assume $\ff(\theta) = 1$  (green) [Equations~\eqref{eq:essick-approx} and \eqref{eq:essick-approx-2}] fail to capture this trend.
    }
    \label{fig:template-bank-holes}
\end{figure}
Search template banks are constructed to target specific fiducial source populations, a choice typically motivated by astrophysical priors. A notable example is the low-mass region of the \textsc{GstLAL} template banks~\citep{2021PhRvD.103h4047M,2021PhRvX..11b1053A,2024PhRvD.109d4066S}, which does not include support for effective spins of $\abs{\chi_{\rm eff}} > 0.05$ at redshifted chirp masses $\mathcal{M}(1+z) \lesssim 1.5 \, M_\odot$. This boundary is adopted because binary neutron stars with such high spins that will merge within a Hubble time have not been observed in the Milky Way~\citep{2009CQGra..26g3001K}. Similar gaps exist in the template banks of other search pipelines~\citep{2017arXiv170501845D,2025CQGra..42j5009A}. Consequently, these searches should exhibit reduced sensitivity to signals occupying such gaps in the parameter space.

To illustrate this effect, we use the BNS injection set used for sensitivity estimates during the LVK's third observing run (O3)~\citep{ligo_scientific_collaboration_and_virgo_2023_7890398, 2023PhRvX..13a1048A}. The O3 injection set has a larger number of BNS injections compared to the O4a injection set, allowing for a stricter test of our prescription. We plot the fraction of detected signals with redshifted chirp masses $\mathcal{M}(1 + z) < 1.5\,M_\odot$, applying a false alarm rate threshold of ${\rm FAR} < 1\,\mathrm{yr}^{-1}$ to the real search results. As shown in Figure~\ref{fig:template-bank-holes}, the search data demonstrates a clear degradation in sensitivity as $\chi_\mathrm{eff}$ deviates from zero.
Our semianalytic prescription successfully reproduces this trend. For our model, we assume a three-detector network (HLV) utilizing the representative O3 PSD from \citet{ligo_scientific_collaboration_2025_16740128}. We adopt the O3 \textsc{GstLAL} template bank~\citep{2021PhRvX..11b1053A} and use an injection distribution identical to that in \citet{2023PhRvX..13a1048A}. To compare our predictions directly with the real search results, we apply a threshold of $\rho_\mathrm{thr}=10$ on \rhonetobs. Our derived detection fractions closely align with the empirical search results and shows that sensitivity degrades more severely for positive $\chi_{\rm eff}$ than for negative $\chi_{\rm eff}$. This asymmetry is primarily driven by the orbital hangup effect~\citep{Campanelli2006}: systems with positive $\chi_\mathrm{eff}$ inspiral for a larger number of cycles, and vice versa for negative $\chi_\mathrm{eff}$ systems. Because the template bank lacks templates at high positive spins, the search algorithm attempts to recover these signals using templates with lower chirp masses to compensate for the extra cycles. However, this compensation is imperfect due differing dependence of the dominant terms involving $\mathcal{M}$ and $\chi_\mathrm{eff}$ in the post-Newtonian expansion, and the ability to shift to a lower chirp mass is strictly bounded by the low-mass edge of the template bank. In contrast, the bank extends to much higher chirp masses, providing ample templates to absorb the shorter waveforms of negative $\chi_\mathrm{eff}$ systems. This explains the comparatively smaller drop in sensitivity for negative spins. The sensitivity drops by a factor of ${\sim} 2$ at $\chi_{\rm eff} \approx 0.2$, worsening to a factor of $>3$ for systems reaching $\chi_{\rm eff} \approx 0.3$.

During the LVK's fourth observing run (O4), no un-retracted public alerts~\citep{gracedb} were issued for BNSs. Consequently, the estimated BNS merger rates have been revised downward in the latest catalog when compared to the estimates following the detection of GW170817~\citep{2025arXiv250818083T}. While there are a number of well-motivated astrophysical explanations for this dearth of detections~\citep{Fishbach2026, Kunnumkai2026}, one speculative proposal in the context of the result in Figure~\ref{fig:template-bank-holes} is that there exists a significant population of BNSs at high spin that the template bank considered here is not adequately equipped to detect\footnote{Other flavors of searches such as the \textsc{GstLAL} search for sub-solar mass objects~\citep{2025PhRvD.112d4013H} could be sensitive to high spin configurations.}. 

\begin{figure}[!tbp]
    \centering
    \includegraphics[width=\columnwidth]{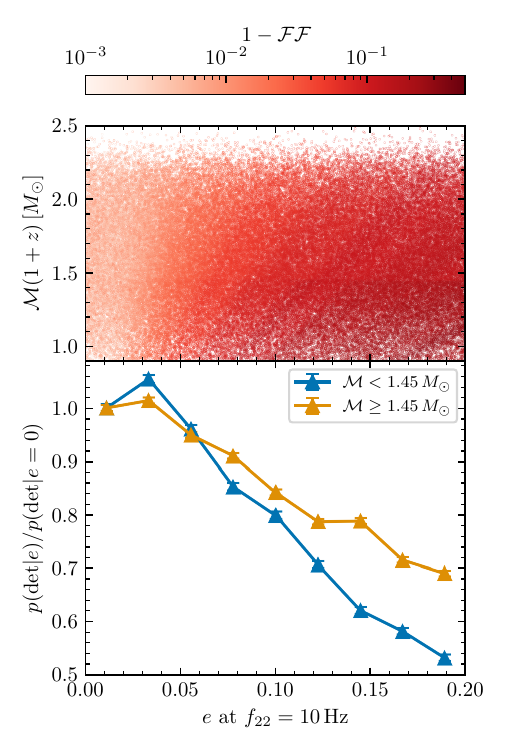}
    \caption{
        \textit{Semianalytic sensitivity estimates for nonspinning, eccentric BNS systems}.
        The top panel shows values of $1 -\ff$ in the $\mathcal{M}(1+z)$-$e$ space, where the eccentricity $e$ is defined at dominant (2,2) mode frequency of $10\,\mathrm{Hz}$.
        The \ff values decrease with increasing $e$.
        In the bottom panel, we plot the ratio $p(\mathrm{det}|e) / p(\mathrm{det} | e=0)$ in different eccentricity bins for two different BNS mass bins. 
    }
    \label{fig:bns-eccentricity}
\end{figure}


\subsection{Known Missing Physics}
\label{sec:known missing physics}
Detecting imprints of orbital eccentricity in GW signals is one of the big goals for GW astronomy.
Apart from having rich features from a phenomenological standpoint, eccentric mergers are also excellent tracers of the progenitor environments of compact binaries.
GW emission is very efficient at radiating away eccentricity from a binary system~\citep{Peters1964}.
Thus, to retain any semblance of eccentricity in the band of the LVK detectors, the binary should have formed at low separations but with high eccentricities.
Dense star clusters~\citep[e.g.,][]{2016ApJ...831..187A, 
2018PhRvL.120o1101R,2019MNRAS.488...47F,2018ApJ...860....5G,2020MNRAS.492.2936A,  2024A&A...683A.186D}, hierarchical triple systems~\citep[e.g.,][]{2003ApJ...598..419W, 2017ApJ...836...39S,2017ApJ...841...77A, 2026MNRAS.545f1938D}, disks of active galactic nuclei~\citep[e.g.,][]{2019PhRvL.123r1101Y,2022Natur.603..237S}, etc., provide natural settings to host such mergers.
Spurred by their astrophysical significance, there have been a number of efforts in recent years to model the waveforms of eccentric signals~\citep{2025PhRvD.112d4038G, 2025PhRvD.111h4052M,2025PhRvD.112l1503A,2026PhRvD.113b4006P,2026PhRvD.113h3044R}.

\begin{figure}[!tbp]
    \centering
    \includegraphics[width=\columnwidth]{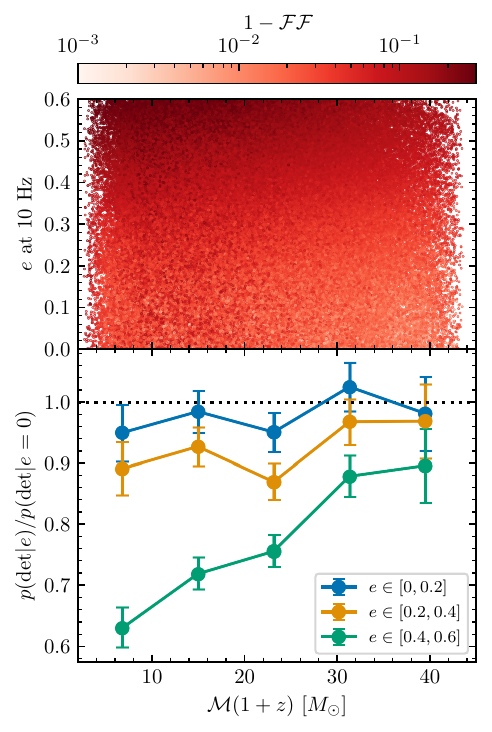}
    \caption{
        \textit{Semianalytic sensitivity estimates for aligned-spin, eccentric BBH systems}.
        The top panel shows the recovered $1 - \ff$ values in the $e$-$\mathcal{M}(1+z)$ plane for an injection set statistically identical to the one used in \citet{2024PhRvD.110d4013G}. 
        The bottom panel shows the the ratio $p(\mathrm{det}|e) / p(\mathrm{det} | e=0)$ for different eccentricity ranges and in different bins of $\mathcal{M}(1+z)$.
    }
    \label{fig:bbh-eccentricity}
\end{figure}

\begin{figure*}[!tbp]
    \centering
    \includegraphics[width=\columnwidth]{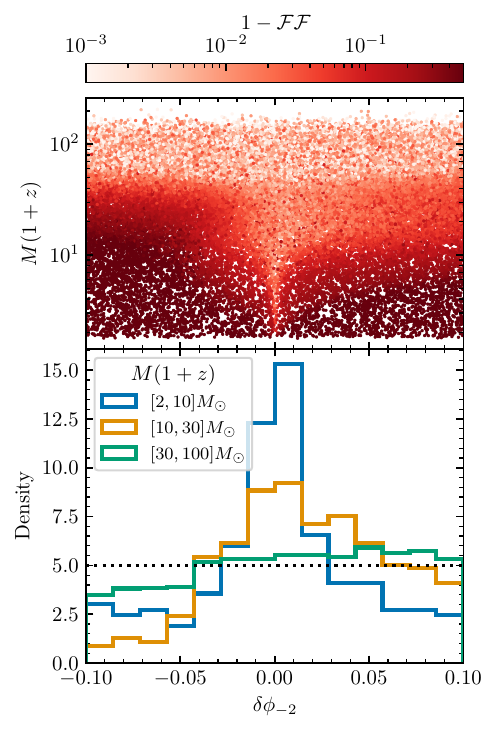}
    \includegraphics[width=\columnwidth]{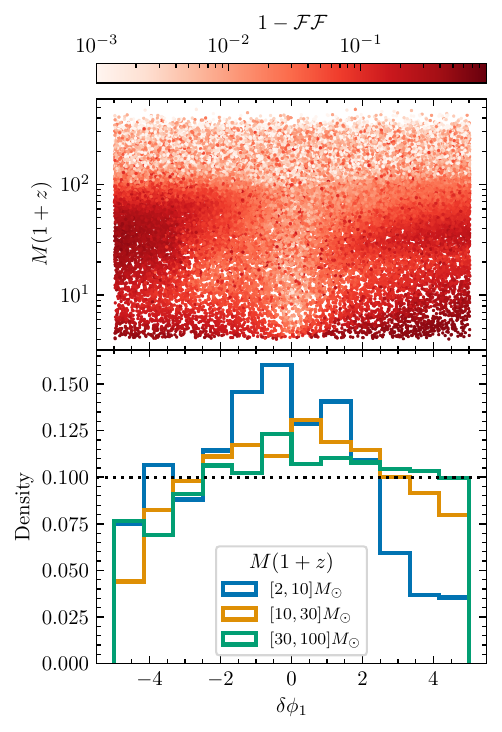}
    \caption{
        \textit{Semianalytic sensitivity estimates for BBH systems with deviations away from general relativity}. The top panels show the values of $1-\ff$ in the $M(1+z)$-$\delta\phi$ parameter space. The bottom panels show the detected distributions for the $\delta \phi_{-2}$ and $\delta \phi_1$ deviation parameters in three different mass bins.
    }
    \label{fig:bbh-tgr}
\end{figure*}

We illustrate the selection function for nonspinning, eccentric BNSs by creating injection set using the \textsc{TaylorF2Ecc} waveform approximant~\citep{2016PhRvD..93l4061M} over a narrow range of eccentricities $0$-$0.2$.
BNS signals are long and hence are amenable to measurements of smaller values of orbital eccentricity as compared to BBHs~\citep{2021ApJ...921L..43Z, 2024ApJ...969..132V}.
This same fact also means that the detectability of eccentric signals drops sharply for nonzero eccentricities, when recovered with a template bank that contains circular signals.
This is also evident from Figure~\ref{fig:bns-eccentricity}, where we see around a factor of ${\sim} 20\%$ drop in sensitivity at $e\approx0.1$ and ${\sim} 30$-$50$\% drop in sensitivity at $e\approx0.2$ depending on the chirp mass range.
Thus, if BNS signals have a small residual eccentricity due to being part of a hierarchical triple system~\citep{Hamers2019} or being produced by through specific natal kick configurations~\citep{Beniamini2024}, a significant fraction will be missed by the current LVK search pipelines.
Thus, searches with orbital eccentricity~\citep{2021ApJ...915...54N, 2025PhRvD.111j3018D, 2025PhRvD.112l2007K, 2026PhRvD.113j3023P} are timely.

To test for sensitivity to eccentric BBHs, we use results from \citet{2024PhRvD.110d4013G} as a point of comparison. 
\citet{2024PhRvD.110d4013G} examine the sensitivity of the \textsc{PyCBC} search pipeline in O3 to eccentric signals.
For this purpose, they create an injection set with a  range of masses between $\mathcal{M}(1 + z)  \in [5, 70] \, M_\odot$ and eccentricities between $0$-$0.6$ using the \textsc{SEOBNRv5EHM} waveform approximant~\citep{2025PhRvD.112d4038G}.
Luminosity distances are distributed uniformly in chirp distance i.e., $D_L [\mathcal{M}(1+z) / M_\odot]^{5/6}$, between $5\,\mathrm{Mpc}$ and $300\,\mathrm{Mpc}$. 
The eccentricity is defined at a reference 22-mode GW frequency of $10\,{\rm Hz}$.
We create a statistically identical injection set (albeit with a larger number of injections), and calculate \rhonetobs assuming the \textsc{PyCBC} template bank and PSDs that were used for the search. 
In Figure~\ref{fig:bbh-eccentricity}, we plot the ratio $p(\mathrm{det}|e) / p(\mathrm{det}| e=0)$ as a function of the injection chirp mass, where the ratio quantifies the probability in a certain bin of eccentricity divided by the detection probability for circular signals under the same distributions for the other source parameters.
We are able to reproduce the 5\%-30\% drops in sensitivity observed across eccentricities for a range of masses. 


\subsection{Unknown Missing Physics}
\label{sec:unknown missing physics}

Yet another example of missing physics in template banks is that of effects beyond general relativity.
Understanding the selection effects inherent for signals from theories beyond general relativity is important when combining information from multiple events in hierarchical tests of general relativity~\citep{2019PhRvL.123l1101I}.

For the purposes of this subsection, we follow \citet{2024PhRvD.109b3014M} and use the results therein as a point of comparison. 
The inspiral dynamics of a compact binary merger can be approximated by post Newtonian (PN) theory. 
Following post-Newtonian theory, the GW phase can be expanded as a series in orbital velocity $v $.
That is, the GW phase in general relativity can be written as,
\begin{equation}
    \psi(v) = -\dfrac{3}{128\eta} v^{-5}\sum_k \phi_k v^k
\end{equation}
where $\phi_k$ is the PN coefficient corresponding to order $k$.
In general relativity, $k$ can take values $0, 2,3,4,\ldots$. One can add \textit{fractional} deviations by parameterizing $\phi_k \rightarrow \phi_k(1 + \delta\phi_k)$ for these orders, and absolute deviations $\delta \phi_k$ for orders that do not appear in general relativity.
Here, we will only consider the deviations at orders $k=-2$ and $k=1$.

We use exactly the same population as in \citet{2024PhRvD.109b3014M} and plot the histogram of detected injections, assuming the O3 \textsc{GstLAL} template banks and PSDs.
The histogram in \citet{2024PhRvD.109b3014M} showed fewer detected signals at negative values of $\delta \phi_{-2}$ compared to positive values.
This asymmetry arises because negative $\delta \phi_{-2}$ values correspond to longer signals, leading to more waveform cycles; as a result, the template banks is efficient at recovering such signals.
This trend is also borne out in our semianalytic sensitivity estimates plotted in Figure~\ref{fig:bbh-tgr}.
There are also strong trends as a function of total mass with greater loss in detections at extreme values of $\delta \phi_{-2}$. Similar trends are visible for $\delta\phi_1$. 
This loss in sensitivity may be ameliorated by designing searches to detect beyond-GR deviations~\citep[e.g.,][]{2023PhRvD.107b4017N, 2024PhRvD.109l4049S}. 


\section{Discussion}
\label{sec:discussion}

There are a number of other approaches in the literature aimed at fast evaluation of selection effects.
One class of such methods abandons physical understanding of the selection function and attempts to fit a (over)parametrized model directly from injections using machine learning techniques~\citep{2020PhRvD.102j3020G, 2020arXiv201201317T,2024PhRvD.110l3041C}.
While these approaches can capture selection biases arising from unmodeled effects in the template bank, they require that injections be processed through a full search pipeline, and thus inherit the same computational and person-power challenges as real search analyses.
They also will not necessarily generalize well, manifestly cannot capture effects that were not in the original injection set, and have many free parameters that often have no clear interpretation.
Other methods that calibrate injections processed by searches to expressions that are simple to evaluate~\citep{2024PhRvD.109f3013M,2025CQGra..42d5008L} suffer from the same issues. 

Another set of methods~\citep{1993PhRvD..47.2198F, Wysocki2019,2024CQGra..41l5002G} use physical principles behind the signal model and detection process, similar in principle to \citet{2023PhRvD.108d3011E}.
These are expected to generalize better and some have relatively few free parameters (e.g., \citealt{2023PhRvD.108d3011E} has only a single free parameters: $\rho_\mathrm{thr}$).
While these approaches may be able to capture effects not originally included in injection sets processed by searches, they assume that the template bank is dense and complete.
Therefore, they cannot account for any of the effects considered in \S\ref{sec:applications}.
The methods developed in this work alleviate both these issues and enable fast evaluation of the selection function for a large class of signals. The injection set thus generated could also be used to train neural network emulators~\citep{2024PhRvD.110l3041C}, particularly for use in simulation-based inference frameworks~\citep{2026arXiv260511274L}.

We emphasize that the semianalytic sensitivity estimates presented here do not replicate the full behaviour of a search pipeline; rather, they approximate only the matched-filtering operation.
Real searches apply a suite of signal consistency tests to construct their detection statistics, like \textsc{PyCBC}'s newsnr~\citep{2013PhRvD..87b4033B} and \textsc{GstLAL}'s autocorrelation $\xi^2$~\citep{2023PhRvD.108d3004T} to verify the consistency of signal power across frequency bins.
These tests are designed to suppress instrumental artefacts and are not expected to affect genuine compact binary signals.
Consistent with this expectation, our results indicate that signal-consistency tests have only a minor impact on injection-based sensitivity estimates and that sensitivity is primarily driven by SNR loss from template mismatch. 
Nonetheless, signals that are highly eccentric or that deviate substantially from the GR waveform manifold may fail $\chi^2$-like consistency tests.
Consequently, our semianalytic estimates should be interpreted as upper limits on the achievable search sensitivity.
Similar conclusions were reported by \citet{2024PhRvD.109b3014M} in their analysis of signals beyond general relativity using real search pipelines. 

Throughout this work, we have assumed that the $\ff(\theta)$ value and the corresponding best match template are the same across detectors.
This may not be strictly true, since detectors that have different PSD shapes would give different values of $\ff(\theta)$. Additionally, we have assumed fixed representative PSDs for ease of computation, but this condition can also be relaxed.
The results of \S\ref{sec:applications} suggest that these assumptions are unlikely to introduce a significant bias, although a more general version of the prescription used in this paper can be incorporated in future extensions.

Our framework identifies the best-match template by maximizing the match over template parameters before projecting onto a detector network and introducing the effects of noise. This contrasts with actual search pipelines, which operate on the SNR of the noisy data across the template bank. However, this simplification is well-justified; as demonstrated by \citet{2023PhRvD.108d3011E}, the strong correlations present across the template bank ensure that this approximation yields reliable estimates for $\rhonetobs$ and the overall detectability. Additionally, we find that the inherent discreteness of template banks means this noise-free maximization approach also serves as a reasonable approximation for recovering the best-match template parameter.

This work only considers quadrupolar aligned-spin template banks, consistent with the template banks employed in the searches we reference. This assumption also simplifies the computation of $\ff(\theta)$.
There is, however, no fundamental obstacle to extending our framework to searches such as the \textsc{IAS-HM} search~\citep{2024PhRvD.110d4063W}.
In that case, the data are independently filtered with templates corresponding to different harmonics, and the resulting detection statistics are subsequently combined in post-processing. 
A similar strategy could be implemented within our semianalytic prescription, although doing so would increase the computational cost by approximately a factor equal to the number of harmonics considered.
Similar prescriptions can be applied to searches that model eccentricity and precession in their template banks.

Our prescription can be used to rapidly identify sensitivity gaps across the parameter to specific class of signals, and can be extended to a variety of applications including gravitational lensing~\citep{2025PhRvD.111h4019C}, astrophysics-induced modulation of the signal~\citep[e.g.,][]{2023ApJ...954..105V, 2025PhRvD.111h4037R, 2025PhRvD.112h4034T}, and a wide variety of effects beyond Kerr and general relativity.  

\section*{Acknowledgements}
We thank Shio Sakon for many useful discussions about  \textsc{GstLAL} template banks and analysis choices, and for a careful reading of our draft. We also thank Kanchan Soni and Bhooshan Gadre for clarifications about \cite{2024PhRvD.110d4013G} and sharing the template bank used in that work. We are grateful to Bart Ripperda for graciously granting us access to the GPU-enabled \textsc{bee} workstation at CITA. This research has made use of the Astrophysics Data System, funded by NASA under Cooperative Agreement 80NSSC21M00561.

This research was supported by the Natural Sciences and Engineering Research Council of Canada (NSERC).
R.E. is supported by NSERC Grant RGPIN-2023-03346, and A.V. is supported by NSERC Grant DIS-2022-568580.
We acknowledge the use of the following software packages: \texttt{gw-detectors}~\citep{gw-detectors}, \texttt{gw-distributions}~\citep{gw-distributions}, \texttt{astropy} \citep{astropy:2013, astropy:2018, astropy:2022}, \texttt{Jupyter} \citep{2007CSE.....9c..21P, kluyver2016jupyter}, \texttt{matplotlib} \citep{Hunter:2007}, \texttt{numpy} \citep{numpy}, \texttt{pandas} \citep{mckinney-proc-scipy-2010, pandas_17229934}, \texttt{python} \citep{python}, \texttt{scipy} \citep{2020SciPy-NMeth, scipy_17101542}, \texttt{Cython} \citep{cython:2011}, \texttt{h5py} \citep{collette_python_hdf5_2014, h5py_7560547}, \texttt{JAX} \citep{jax2018github}, \texttt{LALSuite} \citep{lalsuite, swiglal}, \texttt{seaborn} \citep{Waskom2021}, \texttt{OverCite} \citep{Shariat2026}, and \texttt{tqdm} \citep{tqdm_14231923}.
Part of the software citation information was aggregated using \texttt{\href{https://www.tomwagg.com/software-citation-station/}{The Software Citation Station}} \citep{software-citation-station-paper, software-citation-station-zenodo}.

\appendix


\section{Mass-dependent SNR threshold}
\label{appendix:mass-dependent-rhoth}

We investigate if there is evidence for a mass-dependent SNR threshold ${\rho}_\mathrm{thr}$ when calibrating the semianalytic estimates against real \textsc{GstLAL} results from O4a. In order to find the optimal threshold in each mass bin, we compare the redshift distributions of our semianalytic estimates  under different SNR threshold to injections detected with $\mathrm{FAR} < 1\,\mathrm{yr}^{-1}$ in \textsc{GstLAL}.

Let $g(z)$ denote the redshift distribution from \textsc{GstLAL} and $s(z | \bar{\rho})$ denote the redshift distribution from the semianalytic estimate under a certain SNR threshold $\bar{\rho}$. Let $G(z)$ and $S(z|\bar{\rho})$ be the corresponding cumulative density functions respectively. To quantify the match between distributions, we calculate two different quantities:

\begin{enumerate}
    \item \textbf{Wasserstein distance}: $W\qty(g, s; \bar{\rho}) =  \int \dd{z} \, \abs{G(z) - S\qty(z|\bar{\rho})}$. Also sometimes referred to as the ``earth mover's distance''.
    \item \textbf{Kullback-Leibler (KL) Divergence}: $D_\mathrm{KL}(g, s; \bar{\rho}) = \int{ \dd{z} g(z) \log \dfrac{g(z)}{s\qty(z|\bar{\rho})}} $  
\end{enumerate}

\begin{figure}[!h]
    \centering
    \includegraphics[width=0.5\linewidth]{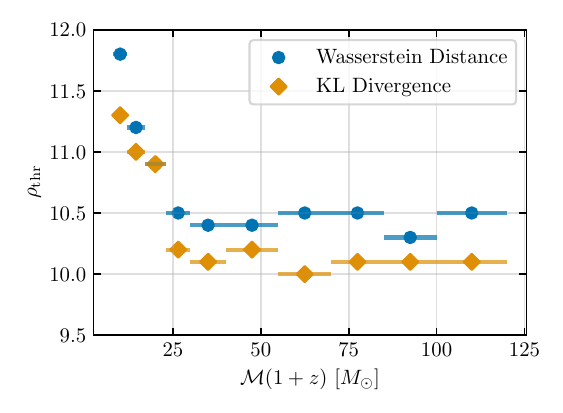}
    \caption{SNR threshold $\rho_\mathrm{thr}$ calibrated to \textsc{GstLAL} results as a function of detector frame chirp mass.}
    \label{fig:mass-dependent-snr-threshold}
\end{figure}

The optimal threshold $\rho_\mathrm{thr}$ is the value of $\bar{\rho}$ that minimizes $W\qty(g, s; \bar{\rho})$ or $D_\mathrm{KL}(g, s;\bar{\rho})$. We plot the derived results in Figure~\ref{fig:mass-dependent-snr-threshold}. Regardless of whether we minimize the Wasserstein distance or the KL divergence, $\rho_\mathrm{thr}$ is constant above $\approx 30 \, M_\odot$ but rises below that value. The $\rho_\mathrm{thr}$ values derived using the KL divergence are systematically lower than the ones derived using the Wasserstein distance. This may be attributed to how these metrics pick up on differences between distributions. The Wasserstein distance is minimized when the bulk of the distribution is similar, since a low probability tail of the distribution doesn't impact the CDF significantly. However, the KL divergence is highly sensitive to such a tail---if $g(z)$ has finite but small support at large redshifts but $s(z|\bar{\rho}) $  has zero support, $D_\mathrm{KL}$ is driven to very high values. A smaller $\bar{\rho}$ places more mergers at high redshift, thus reducing the value of the KL divergence and driving $\rho_\mathrm{thr}$ to smaller values.


\section{Insensitivity of the fitting factor to spacing of the frequency grid}\label{appendix:Delta f}
While calculating the fitting factor, one maximizes the overlap across the relative time offset $t_0^*$ between the candidate signal and the data. 
This time-maximization is performed by taking the inverse Fast Fourier Transform (IFFT) of the noise-weighted inner product of the candidate signal and the data. 
Assuming the minimum and maximum frequencies on the frequency grid are $f_\mathrm{min}$ and $f_\mathrm{max}$, respectively, and the grid spacing is $\Delta f$, the resolution of the IFFT time grid is $(f_\mathrm{max} - f_\mathrm{min})^{-1} \approx 1 \, \mathrm{ms}$, and its length is $1/(\Delta f)$.
For a typical template bank, $t_0^*$ is $\mathcal{O}(10\text{--}100)\,\mathrm{ms}$, which can be very well resolved with the $\approx 1 \mathrm{ms}$ resolution.
Further, because the resolution of the time grid is  independent of the frequency grid spacing, one can choose a large $\Delta f$ and still obtain an accurate match and fitting factor. 
This drastically reduces computational cost, which scales inversely with $\Delta f$. 
Increasing $\Delta f$ naturally introduces a numerical discretization error in the match calculation; however, this error is expected to be quite small, even when using $\Delta f = 4 \, \mathrm{Hz}$.

To test this empirically, we calculated the fitting factors for nonspinning, equal-mass binary black hole mergers with chirp masses between $6\,M_\odot$ and $100\,M_\odot$ using  $\Delta f = 1/16 \, \mathrm{Hz}$ and $\Delta f = 4 \, \mathrm{Hz}$. The maximum absolute difference in the resulting fitting factors was found to be $0.005$, which is negligible for all purposes of this work.


\bibliographystyle{aasjournalv7}
\bibliography{references}


\end{document}